\documentclass[prd,twocolumn,nofootinbib,superscriptaddress,amsmath,amssymb]{revtex4-2}
\usepackage[utf8]{inputenc}
\usepackage{mathtools}
\usepackage{graphics}
\usepackage{graphicx}
\usepackage{dcolumn}
\usepackage{bm}
\usepackage{dsfont} 
\usepackage{amsmath,amssymb}
\usepackage{hyperref}
\usepackage{tabularx}
\usepackage{epstopdf}
\usepackage{tensor}
\usepackage{mathrsfs}
\usepackage[normalem]{ulem}
\usepackage[usenames]{color}
\hypersetup{
    colorlinks=true,
    linkcolor=blue,
    filecolor=magenta,      
    urlcolor=blue,
    citecolor=blue
}
\usepackage[dvipsnames]{xcolor}
\usepackage{mathrsfs}
\usepackage{comment}

\newcommand{\GR}{{\mbox{\tiny GR}}}
\newcommand{\NC}{{\mbox{\tiny NC}}}
\newcommand{\beq}{\begin{equation}}
\newcommand{\eeq}{\end{equation}}

\newcommand{\bs}{\boldsymbol}

\newcommand{\UVA}{Department of Physics, University of Virginia, P.O.~Box 400714, Charlottesville, VA 22904-4714, USA}
\newcommand{\Brown}{Brown Theoretical Physics Center and Department of Physics, Brown University, 182 Hope Street, Providence, Rhode Island, 02903}

\begin{document}

\title{Probing Noncommutative Gravity with Gravitational Wave and \\ Binary Pulsar Observations}

 \author{Leah Jenks}
\affiliation{\Brown}
 \author{Kent Yagi}
 \affiliation{\UVA}
  \author{Stephon Alexander}
\affiliation{\Brown}

\date{\today}

\begin{abstract} 
Noncommutative gravity is a natural method of quantizing spacetime by promoting the spacetime coordinates themselves to operators which do not commute. This approach is motivated, for example, from a quantum gravity perspective, among others. 
Noncommutative gravity has been tested against the binary black hole merger event GW150914. Here, we extend and improve upon such a previous analysis by (i) relaxing an assumption made on the preferred direction due to noncommutativity, (ii) using posterior samples produced by the LIGO/Virgo Collaborations, (iii) consider other gravitational wave events, namely GW151226, GW170608, GW170814 and GW170817, and (iv) consider binary pulsar observations. Using Kepler's law that contains the noncommutative effect at second post-Newtonian order, we derive corrections to the gravitational waveform phase and the pericenter precession. Using the gravitational wave and double pulsar binary observations, we find bounds on a space-time noncommutative tensor $\theta^{0i}$ in terms of the preferred frame direction with respect to the orientation of each binary. We find that the gravitational wave bounds are stronger than the binary pulsar one by an order of magnitude and the noncommutative tensor normalized by the Planck length and time is constrained to be of order unity.
\end{abstract}

\maketitle

\section{Introduction}

Since the advent of gravitational wave astronomy with the detection of gravitational waves (GWs) by the LIGO/Virgo collaboration (LVC), the theory of general relativity (GR) has been directly testable to greater precision than previously possible. Although no observations have yet indicated any compelling deviations from GR, we are able to study modifications to GR, alternative theories of gravity and other fundamental physics using gravitational waves as a probe \cite{TheLIGOScientific:2016src,Yunes:2016jcc,Berti:2018cxi,Berti:2018vdi,LIGOScientific:2019fpa,Abbott:2018lct}. Particularly, non-GR effects are highly constrained by GW observations, which can be used to explore many different theories. This has been done, for example for Einstein-Aether theory \cite{Zhang:2019iim}, Einstein-dilaton-Gauss-Bonet gravity \cite{Nair:2019iur,Yamada:2019zrb}, dynamical Chern-Simons gravity \cite{Nair:2019iur} and others \cite{Carson:2019kkh, Tahura:2018zuq}.

In addition to gravitational waves, pulsar timing observations are valuable tool in probing modifications to GR. The system that we will be using to place constraints via pulsar observations is the double pulsar binary system PSR J0737-3039A/B. This system is quite unique, as both neutron stars are radio pulsars, which allows for extremely precise measurements and provides a rich background for tests of general relativity and modified theories of gravity \cite{Kramer:2006nb,Berti:2015itd}. 

In this paper, we will employ a combination of gravitational wave and pulsar analysis to introduce two independent constraints on noncommutative theories. 
Various noncommutative theories have been proposed previously, originally introduced as a method of quantizing spacetime \cite{PhysRev.71.38}. The introduction of noncommutative geometry \cite{Connes1985} allowed this idea to be applied more broadly, with a focus on noncommutative quantum field theories \cite{Douglas:2001ba, Szabo:2001kg} as well as multiple formulations of a noncommutative extension to the Standard Model \cite{Chaichian:2001py, Chaichian:2004yw, Calmet:2001na, Aschieri:2002mc}. The idea of noncommutative gravity stems from these theories. Non-commuting conjugate variables are a cornerstone of quantum mechanics, and it seems natural that one could apply the same conventions that give rise to, for example, the Heisenberg uncertainty principle in quantum mechanics, to a gravitational setting \cite{Douglas:2001ba, Szabo:2001kg}. Noncommutative gravity also has string theory implications \cite{Ardalan:1998ce, Seiberg:1999vs} and thus we have a wide range of motivations for its study. 
The version that we will be focused on is characterized by promoting spacetime coordinates to operators which satisfy the following canonical commutation relation
\beq 
\label{eq:NC_def}
[\hat{x}^\mu, \hat{x}^\nu] = i\theta^{\mu\nu}.
\eeq 
Here, $\theta^{\mu\nu}$ introduces a new fundamental quantum scale which represents the ``quantum fuzziness" of spacetime, in analogy to $\hbar$ in quantum mechanics.

Previous work \cite{Kobakhidze:2016cqh} has placed a bound on the time component of the noncommutativity scale $\theta^{0i}$ using GW150914, and found a constraint $\sqrt{\Lambda} \lesssim 3.5$, at the order of the Planck scale (see~\cite{Nelson:2010ru,Nelson:2010rt} for related works). The authors worked in the post-Newtonian (PN) formalism, in which quantities are expanded in powers of $(v/c)^n$ with $v$ representing the relative velocity of the binary constituents, which are considered order $n/2$PN \cite{Blanchet:2000cw}. Reference \cite{Kobakhidze:2016cqh} found corrections entering at 2PN in the acceleration and the waveform phase. The authors introduce the notation
\begin{equation}\label{eq:Lambda2}
\Lambda\theta^{i} = \frac{\theta^{0i}}{l_pt_p},
\end{equation}
where $\theta^i$ represent the components of a three-dimensional unit vector $\bs{\theta}$, which acts as a preferred direction that induces precession of the orbital plane. For calculational simplicity, the authors assumed that the orientation of the $\bs{\theta}$  is orthogonal to the orbital plane as to place an approximate upper bound on $\sqrt{\Lambda}$.

In this work we extend and generalize the above analysis by considering the general case for the orientation of the preferred direction $\bs{\theta}$ with respect to the orbital plane by adopting orbital averaging. We then place constraints by employing posterior samples from the GWTC-1 catalog following \cite{Nair:2019iur}, rather than explicitly using the bound on the non-GR parameter at 2PN order found by LVC \cite{TheLIGOScientific:2016src} as done in \cite{Kobakhidze:2016cqh}. This new approach properly accounts for the uncertainties in the masses.We derive bounds from four different gravitational wave events with relatively small masses, namely GW151226, GW170608, GW170814 and GW170817.

We also  place constraints on the the time component of the noncommutative tensor using the binary pulsar system PSR J0737-3039A/B to act as an independent check on the gravitational wave constraints. 
In the binary pulsar system, the noncommutativity induces an additional contribution in the pericenter precession beyond GR at 1PN, due to the preferred direction $\bs{\theta}$ that is induced by the inclusion of noncommutative terms. Corrections to other observables, such as the mass ratio and Shapiro delay, enter at higher PN orders. Thus, we use the latter to determine the masses of the double pulsar binary and use the pericenter precession to constrain the theory (see e.g. \cite{Deng:2017hkj} for a related work on constraining noncommutative gravity from the pericenter precession of binary pulsars). We found that such bounds are slightly weaker than the ones from gravitational wave events.

The structure of the paper is as follows. In Section \ref{II} we derive the lowest order 2PN noncommutative corrections to the binary system acceleration, beginning from the energy-momentum tensor. We proceed to constrain the noncommutative parameter with LVC data in Section \ref{III} by computing the 2PN non-commutative correction to the gravitational waveform. We then use posterior samples for two different waveform templates to constrain the noncommutativity parameter, $\sqrt{\Lambda}$, in terms of the quantity $\bm{\hat{L}}\cdot \boldsymbol{\theta}$. In Section \ref{IV} we then independently constrain $\sqrt{\Lambda}$ as a function of $\bm{\hat{L}}\cdot \boldsymbol{\theta}$ by computing the noncommutative correction to the GR pericenter precession and using the binary pulsar event PSR J0737-3039A/B. Finally, In section \ref{V}, we summarize our results, and provide some concluding remarks as well as directions for future work. We work in the geometric units $c=G=1$.


\section{Noncommutative Corrections to the Acceleration and Energy}\label{II}

In GR, one can approximate a binary system as two point masses which have an energy momentum tensor given by 
\beq \label{eq:T}
T_{\GR}^{\mu\nu}(\bs{x}, t) = m_1\gamma_1(t)v_1^\mu(t)v_1^\nu(t)\delta^3[\bs{x} - \bs{y}_1(t)] + 1\leftrightarrow 2.
\eeq 
Here, $m_i$ are the masses of each body, $\bs{y}_i$ the positions and $v_i^\mu$ the four velocities. $\gamma_i$ is given by 
\beq 
\gamma_i = \frac{1}{\sqrt{g_i(g_{\alpha\beta})_i (v_i^\alpha v_i^\beta/c^2)}}, 
\eeq 
where $g_{\mu\nu}$ is the metric, $g$ its determinant and $i = 1,2$ \cite{Blanchet:2000cw}. It was previously shown in \cite{Kobakhidze:2016cqh} that noncommutative corrections to the expression \ref{eq:T} can be found by considering that the black holes are sourced by a massive real scalar field $\phi$ and incorporating the noncommuting operators $\hat{x}^\mu$ by replacing the product of any two functions with a Moyal product. It was found that the energy-momentum tensor, including noncommutative corrections, can be written as 
\begin{widetext}
\beq \label{eq:Tfull}
\begin{split}
T^{\mu\nu}_{\NC}(\bs{x}, t) = m\gamma_L(t)v^\mu(t)v^\nu(t)\delta^3[\bs{x} - \bs{y}_1(t)] 
+\frac{m^3\gamma_L^3}{8}v^\mu v^\nu \Theta^{kl}\partial_k\partial_l\delta^3[\bs{x} - \bs{y}_1(t)]\\
+ (\eta^{\mu m}\eta^{\nu n}\partial_m\partial_n - \eta^{\mu\nu}\partial_i\partial^i)
 \left(\frac{\hbar^2}{4m\gamma_L} + \frac{m\gamma_L \hbar^2 }{32}\Theta^{kl}\partial_k\partial_l\right)\delta^3[\bs{x} - \bs{y}_1(t)],
\end{split}
\eeq 
\end{widetext}
 where $\gamma_L$ is the Lorentz factor and 
we define 
\beq \label{eq:Theta}
\Theta^{kl} = \frac{\theta^{0k}\theta^{0l}}{l_p^2t_p^2} + 2 v_p\frac{\theta^{0k}\theta^{pl}}{l_p^3t_p} + v_pv_q\frac{\theta^{kp}\theta^{lq}}{l_p^4}.
\eeq 
Here, $\theta$ is the noncommutativity parameter defined by Eq. \eqref{eq:NC_def} while $l_p$ and $t_p$ are the Planck length and time respectively. The second term in Eq. \eqref{eq:Tfull} is suppressed by a factor of $\hbar^2$ and can be neglected. We will consider the contribution from the first term in Eq. \eqref{eq:Theta}, which using the convention that a term of order $(v/c)^n$ is of order $(n/2)$PN, enters as a correction at the second Post-Newtonian order (2PN). We will consider only lowest order noncommutative corrections, and thus can make the approximation $\gamma_L = 1$.
Then, for a binary system which considers only the lowest order noncommutative corrections, the energy-momentum tensor simplifies to  
\beq 
\begin{split}
T^{\mu\nu}_{\NC}(\textbf{x}, t) = m_1\gamma_1(t)v^\mu_1(t)v^\nu_1(t)\delta^3(\textbf{x} - \textbf{y}_1(t)) \\
+ \frac{m_1^3 \Lambda^2}{8}v^\mu_1(t) v^\nu_1(t)\theta^k\theta^l \partial_k\partial_l \delta^3(\textbf{x} - \textbf{y}_1(t)) + 1 \leftrightarrow 2, 
\end{split}
\eeq 
where we have defined a normalization of the noncommutative tensor, $\Lambda$ as in Eq. \ref{eq:Lambda2}.
In analogy to \cite{Kobakhidze:2016cqh}, we follow the standard procedure to arrive at the acceleration, where we consider only the leading order GR contribution and the lowest order noncommutative correction entering at 2PN: 
\beq 
a_i  = (a_i)_{\GR} - \frac{15M^3(1 - 2\nu)\Lambda^2}{8r^4}\theta^k\theta^l\hat{n}_{ikl}.
\eeq 
Here, $M=m_1+m_2$ is the total mass, $\nu = m_1 m_2/M^2$ is the symmetric mass ratio while $r$ is the binary separation.
We have also introduced the quantities $r = |\bs{y}_1 - \bs{y}_2|$ and $\bs{n}$ such that $\bs{n} = (\bs{y}_1 - \bs{y}_2)/r$ to define the quantity 
\beq
\hat{n}_{ikl} =n_in_kn_l - \frac{1}{5}(\delta_{kl} n_i + \delta_{il}n_k + \delta_{ki}n_l).
\eeq
From the acceleration we can also determine the correction to the GR Lagrangian   
\beq 
L = L_{\GR} + \frac{3M^3\mu(1 - 2\nu)\Lambda^2}{8r^3}\theta^k\theta^l\hat{n}_{kl},
\eeq 
and the conserved energy: 
\beq 
E = E_{\GR} - \frac{3M^3\mu(1 - 2\nu)\Lambda^2}{8r^3}\theta^k\theta^l\hat{n}_{kl}. 
\eeq 
Here, $\mu = m_1 m_2/M$ is the reduced mass and
\beq
\hat{n}_{kl} = n_kn_l - \frac{\delta_{kl}}{3}. 
\eeq 

In these expressions for the acceleration, conserved energy and Lagrangian, the vector $\bs{\theta}$ acts as a preferred direction and will in general induce precession in the orbital plane. Previous work \cite{Kobakhidze:2016cqh} simplified these expressions for the acceleration, Lagrangian and conserved energy by assuming a constrained case in which the orbital plane is perpendicular to the preferred direction, $\bs{\theta}$. Given that each binary is expected to be oriented randomly with respect to the preferred direction, the chance of the above assumption being satisfied seems extremely low. To overcome this, we perform an orbital averaging procedure as is typically done for precessing  \cite{Kidder:1992fr,Kidder:1995zr} and magnetized \cite{Ioka:2000yb} binaries. We will consider the following relation as an orbital average over the unit vector $\bs{n}$ and the preferred direction $\bs{\theta}$ as follows:
\beq 
\overline{(\textbf{n}\cdot\boldsymbol{\theta})(\textbf{n}\cdot\boldsymbol{\theta}) }= \frac{1}{2}\left(1 - (\bm{\hat{L}}\cdot \boldsymbol{\theta})^2\right). 
\eeq
Here, $\hat{\bs{L}}$ is a unit vector orthogonal to the orbital plane, as the projection of the angular momentum of the binary system. The case $\bm{\hat{L}}\cdot \boldsymbol{\theta} =1$ corresponds to the limiting case in which the preferred direction is perpendicular to the orbital plane. Employing the orbital averaging procedure, we obtain for the acceleration and conserved energy: 
\begin{align}
\label{eq:aTot}
a_i =& (a_i)_{\GR}- \frac{15M^3(1-2\nu) \Lambda^2}{8r^4} \nonumber \\
&\times \left( n_i (\textbf{n}\cdot\boldsymbol{\theta})^2 - \frac{1}{5}n_i - \frac{2}{5}\theta_i(\textbf{n}\cdot \boldsymbol{\theta})\right),
\end{align}
and
\beq 
E = E_{\GR} - \frac{M^3\mu(1 - 2\nu)\Lambda^2}{16r^3}\left( 1 - 3(\bm{\hat{L}}\cdot \boldsymbol{\theta})^2\right). 
\eeq 


\section{Gravitational Wave Constraints}\label{III}

In this section, we study bounds on noncommutative gravity with gravitational wave observations. We first derive corrections to the gravitational waveform phase. We then find bounds on $\sqrt{\Lambda}$ using posterior samples of selected gravitational wave events produced by LVC.

\subsection{Gravitational Waveform}
From the acceleration and the conserved energy, we can compute noncommutative corrections to the gravitational waveform to constrain the theory. We focus on a quasicircular orbit such that $r$ is a constant. Defining the relative position as $\textbf{y}(t) = \textbf{y}_1(t) - \textbf{y}_2(t)$, we can rewrite Eq. \eqref{eq:aTot} as
\beq 
\textbf{a} = -\Omega^2\textbf{y} + \mathcal{O}(1/c^5). 
\eeq 
In order to find the leading noncommutatice correction to the waveform, we here keep only the leading GR and 2PN NC term. The angular velocity $\Omega$ is given by 
\beq 
\label{eq:Omega2}
\Omega^2 = \frac{M}{r^3}\left[1 + \frac{3(1-2\nu)\Lambda^2}{16}(1 - 3 (\bm{\hat{L}}\cdot \boldsymbol{\theta})^2)\gamma^2\right], 
\eeq
where we have defined the quantity
\beq
\gamma = \frac{M}{r}.
\eeq 
Similarly, taking into account both the explicit 2PN contribution to the energy as well as the 2PN correction to $\Omega^2$ in the leading order GR contribution, we have 
\beq 
E = -\frac{\mu \gamma}{2}\left[1 - \frac{1}{16}(1-2\nu)\Lambda^2\left( 1 - 3(\bm{\hat{L}}\cdot \boldsymbol{\theta})^2\right)\gamma^2\right].
\eeq 
Then, inverting Eq. \eqref{eq:Omega2}  and defining the quantity $
x = \left(M\Omega\right)^{2/3} $ that corresponds to relative velocity squared, we can rewrite the conserved energy in terms of $x$:
\beq 
E = -\frac{\mu x}{2}\left[ 1 - \frac{1}{8}(1-2\nu)\Lambda^2\left( 1 - 3(\bm{\hat{L}}\cdot \boldsymbol{\theta})^2\right)  x^2\right].
\eeq 

To determine the lowest order noncommutative corrections to the energy radiated by gravitational waves, we assume the energy balance equation 
\beq 
\frac{dE}{dt} = -\mathcal{L}.
\eeq 
Here, $\mathcal{L}$ is the gravitational wave luminosity, which is given by 
\beq \label{eq:Fdef}
\mathcal{L} = \left[\frac{1}{5}\frac{d^3 I_{ij}}{dt^3}\frac{d^3I_{ij}}{dt^3} + \mathcal{O}(1/c^2)\right],
\eeq 
where $I_{ij}$ is the traceless mass quadrupole moment. There are two noncommutative corrections to the quadrupole moment. The explicit 2PN noncommutative contribution is time independent, as shown in \cite{Kobakhidze:2016cqh}, and will not contribute to the gravitational wave luminosity. Thus, we only need to consider the Newtonian part of $I_{ij}$, which will lead to noncommutative corrections through the acceleration. For the leading order and 2PN noncommutative corrections to the third derivative of the quadrupole moment, we find
\beq 
\begin{split}
\dddot{I}_{ij} =  - \frac{8\nu M^2}{r^3}\left(\frac{y_iv_j + v_iy_j}{2}\right)\\
\times\Bigg[1 + \frac{15}{8}\Lambda^2(1-2\nu)\left( (\bs{n}\cdot\bs{\theta})^2 - \frac{1}{5}\right)\gamma^2\Bigg ]\\
+ \frac{9\nu M^4}{4 r^4}\Lambda^2(1-2\nu)(\bs{n}\cdot\bs{\theta})(\theta_iv_j + \theta_jv_i).
\end{split}
\eeq 
Squaring and keeping only the relevant lowest order terms, we find after orbital averaging and inserting the result into Eq. \eqref{eq:Fdef} that the full expression for the luminosity is
\beq \label{eq:flux}
\mathcal{L} = \frac{32}{5}\nu^2x^5\left[1 + \frac{\Lambda^2(1-2\nu) }{32}\left(23 - 39 (\hat{\bs{L}}\cdot \bs{\theta})^2 \right)x^2\right].
\eeq 

It is then straightforward to determine the evolution of the orbital phase of the binary system. We define a new parameter 
\beq 
\Theta \equiv \frac{\nu}{5GM}(t_c - t),
\eeq
where $t_c$ is the coalescence time, such that the energy balance equation can be written as 
\beq 
\frac{dE}{dx}\frac{dx}{d\Theta} = \frac{5M}{\nu}\mathcal{L}. 
\eeq 
This can then be solved order by order to find 
\beq 
x = \frac{1}{4}\Theta^{-1/4}\left\{ 1 -  \frac{\Lambda^2(1-2\nu)}{1024}[35 -75 (\hat{\bs{L}}\cdot \bs{\theta})^2] \Theta^{-1/2}\right\}. 
\eeq 
We then invert this expression for $x$ to find  $\Theta$, 
\beq \label{eq:Theta}
\Theta = \frac{1}{256x^{4}}\left\{1 - \frac{\Lambda^2(1-2\nu)}{16}\left[35 - 75 (\hat{\bs{L}}\cdot \bs{\theta})^2\right]x^2\right\}. 
\eeq 
$\Theta$ is related to the orbital phase by the following
\beq 
\frac{d\phi}{d\Theta} = -\frac{5}{\nu}x^{3/2}, 
\eeq 
which can easily be solved to find 
\beq \label{eq:phi}
\phi = -\frac{x^{-5/2}}{32\nu}\left\{ 1 -\frac{5}{32}\Lambda^2(1-2\nu)\left[35 - 75 (\hat{\bs{L}}\cdot \bs{\theta})^2\right]x^2\right\}. 
\eeq

Then, as we have assumed that the velocity of each binary component is small compared to $c$, we may use the  stationary phase approximation (SPA) \cite{Yagi:2013du}, under which the phase of the waveform in Fourier domain can be written as 
\beq 
\psi(f) = 2\pi f t_f - \frac{\pi}{4} - \Phi(t_f). 
\eeq 
$t_f$ is the time such that $d\Phi(t_f)/dt = f$. It can be found from \eqref{eq:Theta}, and $\Phi(t_f)$ is found from \eqref{eq:phi} to obtain the full expression for the inspiral phase including the leading order term and explicit 2PN noncommutative correction:
\beq 
\begin{split}
\psi_I(f) = 2\pi ft_c - \phi_c -\frac{\pi}{4} + \frac{3}{128\nu}\left(\pi Mf\right)^{-5/3}\\
\times\left\{1 - \frac{5}{16}\Lambda^2(1-2\nu)\left[35 -75 (\hat{\bs{L}}\cdot \bs{\theta})^2\right]\left(\pi Mf\right)^{4/3}\right\}. 
\end{split}
\eeq 
 This expression follows the standard PN waveform format, 
 \beq 
\psi_I(f) = 2\pi ft_c - \phi_c - \frac{\pi}{4} + \frac{3}{128\nu}\sum_{j=0}^4 \varphi_j \left(\pi M f\right)^{(j-5/3)}. 
\eeq 

We will be interested in the $\varphi_4$ coefficient, which is what enters at 2PN. Including our correction in addition to the 2PN GR contribution to $\varphi_4$ \cite{Damour:2000zb}, we have
\begin{align}
\varphi_4 = &\frac{15293365}{508032} + \frac{27145}{504}\nu + \frac{3085}{72}\nu^2 \nonumber \\
&- \frac{5}{16}\Lambda^2(1-2\nu)\left[35 - 75 (\hat{\bs{L}}\cdot \bs{\theta})^2\right].
\end{align} 
We can then define the fractional deviation from GR as
\begin{align}
\label{eq:deltaphi4}
\delta\varphi_4^{\NC} \equiv & \frac{\varphi_4^{\NC}}{\varphi_4^{\GR}}\nonumber \\ =&\frac{158760 (1-2\nu)}{4353552 \nu^2 + 5472432 \nu + 3058673} \nonumber \\
&\times\left[-7 + 15 (\hat{\bs{L}}\cdot\bs{\theta})^2\right]\Lambda^2.
\end{align} 
We can now employ this result to constrain the quantity $\left[-7 + 15 (\hat{\bs{L}}\cdot\bs{\theta})^2\right]\Lambda^2$ from gravitational wave events. 

\subsection{Bounds on $\sqrt{\Lambda}$ }

Having obtained the expression for the noncommutative correction to $\varphi_4$, it is now straightforward to compute bounds. However, one issue we still face is the presence of $\nu$, the symmetric mass ratio in the expression $\delta\varphi_4^{NC}$. One could simply take the central values given for each binary component mass to compute $\nu$, however this method does not take into account the uncertainties in the mass and will not give as precise of an answer. As an alternative, we will follow the method outlined in \cite{Nair:2019iur} and make use of the LVC posterior samples for multiple events in order to obtain 90\% confidence bounds on the noncommutative parameter. 

We use posterior samples from the GWTC-1 catalog for gravitational wave events GW 151226, GW170608 and GW170814 \cite{LIGOScientific:2018mvr, LIGOScientific:2019fpa}. Data for events GW150914 and GW170104 are also available, however these two events are characterized by large masses and thus a short inspiral period. This makes it difficult to reliably probe non-GR effects through corrections to the waveform~\cite{TheLIGOScientific:2016src}, thus we do not include constraints from these events. We do however also calculate constraints based on the binary neutron star event GW170817, for which posterior samples are also available \cite{TheLIGOScientific:2017qsa, Abbott:2018lct}.

\begin{figure*}
    \centering
    \includegraphics[width=8.5cm]{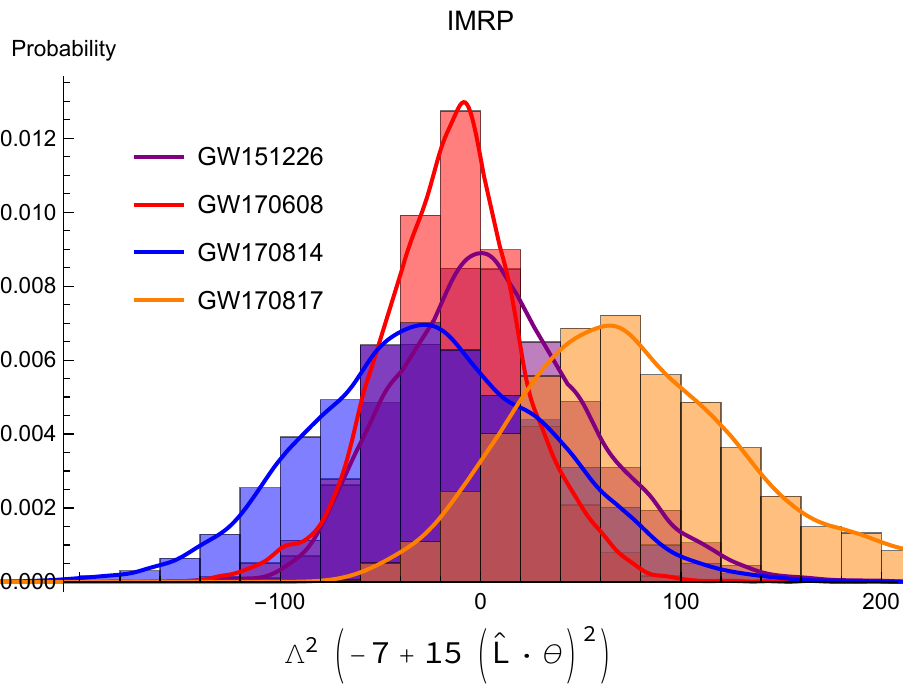}
 \includegraphics[width=8.5cm]{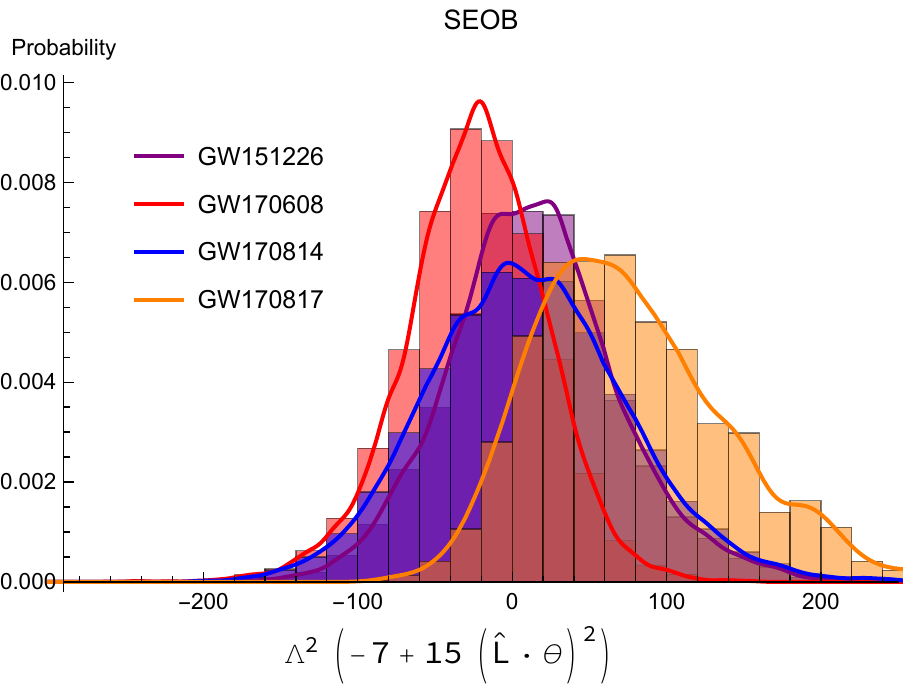}   \caption{Posterior distributions of $\Lambda^2(-7 + 15 (\hat{\bs{L}}\cdot\bs{\theta})^2)$ for various gravitational wave events derived from the posterior samples using the IMRP (left) and SEOB (right) waveform templates.}
    \label{fig:histogram}
\end{figure*}

Inverting Eq. \eqref{eq:deltaphi4} allows us to obtain an expression for $\left(-7 + 15 (\hat{\bs{L}}\cdot\bs{\theta})^2\right)\Lambda^2$ in terms of $\delta\varphi_4$, $m_1$ and $m_2$. 
Then, using posterior samples for the two waveform templates IMRPhenomPv2 (IMRP) and SEOBNRv4 (SEOB), we are able to plot the histograms and probability distribution functions (PDFs) for each event, shown in Fig. \ref{fig:histogram}. 

\begin{figure}
    \centering
    \includegraphics[width=\linewidth]{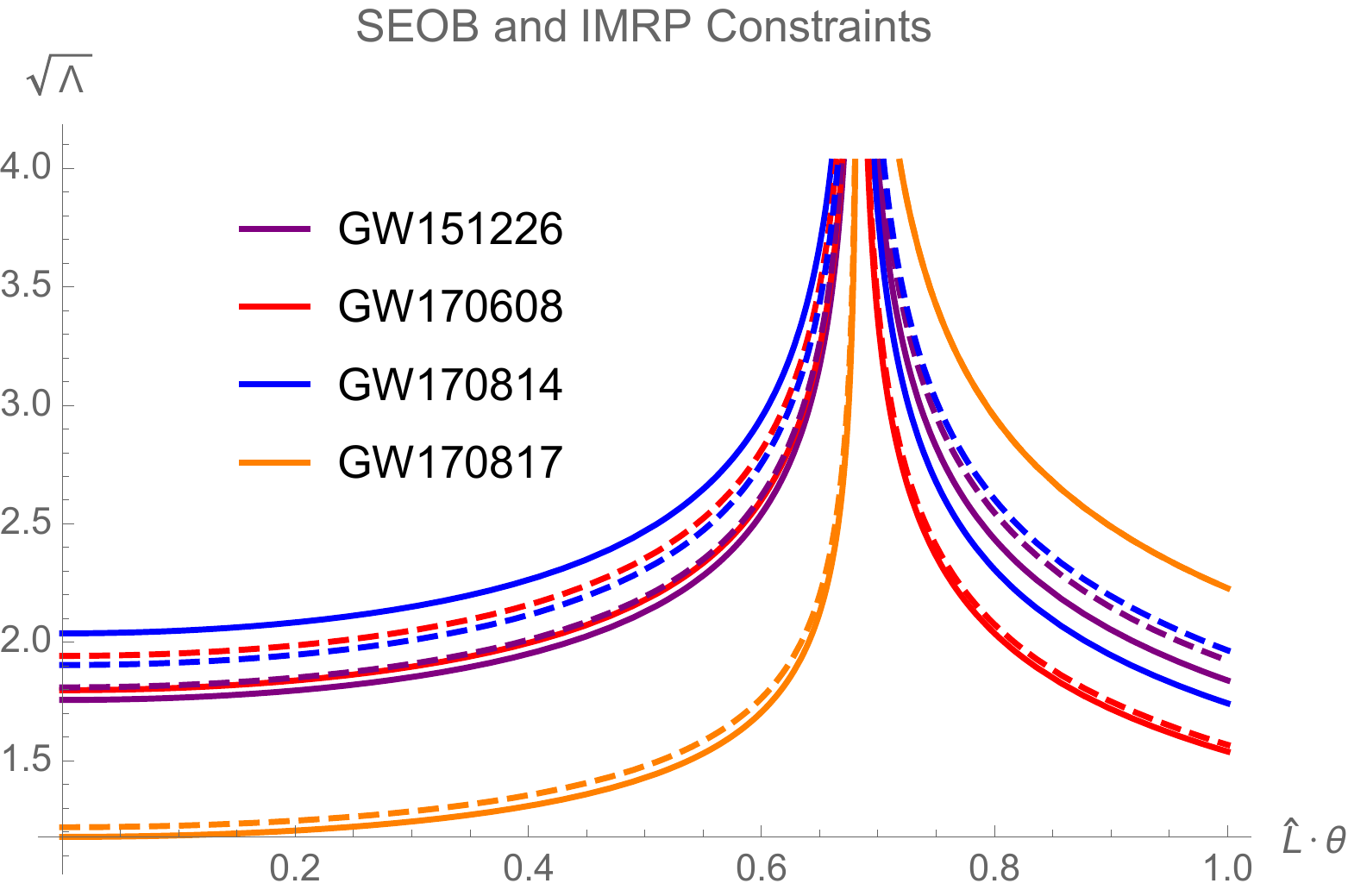}
    \caption{Constraints on the noncommutative parameter $\Lambda$ for each gravitational wave event from SEOB (dashed) and IMRP (solid) waveform templates.  
    }
    \label{fig:GW_bound}
\end{figure}

From the PDFs for each event, we calculate 90\% constraints on $\Lambda^2[-7 + 15 (\hat{\bs{L}}\cdot\bs{\theta})^2]$ as an upper and lower bound. We can then use these upper and lower bounds to constrain $\sqrt{\Lambda}$ as a function of $\hat{\bs{L}}\cdot\bs{\theta}$. These constraints are shown in Fig. \ref{fig:GW_bound} for both waveform templates.
From Fig. \ref{fig:GW_bound}, we see that there is a region of the $\hat{\bs{L}}\cdot\bs{\theta}$ plane in which we cannot constrain the noncommutativity parameter, specifically when $\hat{\bs{L}}\cdot\bs{\theta} = \sqrt{7/15}$. However, given that we are considering multiple gravitational wave events and $\hat{\bs{L}}\cdot\bs{\theta}$ varies from one binary to another, statistically the chance that each of those events would be specifically at $ \hat{\bs{L}}\cdot\bs{\theta} = \sqrt{7/15}$ is low, thus we expect we can still place meaningful bounds.  In total, we can see that the 90\% confidence constraints on $\sqrt{\Lambda}$ as a function of  $\hat{\bs{L}}\cdot\bs{\theta}$ is indeed constrained to be of order unity, as was previously found in \cite{Kobakhidze:2016cqh}.

\section{Binary Pulsar Constraints} \label{IV}
We now turn to constraints on the noncommutativity tensor from the double pulsar binary system PSR J0737-3039A/B \cite{Kramer:2006nb}. We first derive noncommutative corrections to the pericenter precession. We then find bounds on $\sqrt{\Lambda}$ using the double pulsar system.

\subsection{Pericenter Precession}

Beginning from the acceleration, Eq. \eqref{eq:aTot}, we can easily calculate the correction to the  pericenter precession to provide another independent bound on $\sqrt{\Lambda}$. We will treat the noncommutative correction to the acceleration as a perturbing acceleration $\delta\bs{a}$. We can then define the orbital parameters following a standard formulation of osculating orbits explained e.g. in \cite{Will:2018bme} and find the correction to the pericenter precession, given by
\beq
\begin{split} \label{eq:dwdt}
 \frac{d\omega}{dt} =\frac{1}{e}\sqrt{\frac{p}{M}}\Bigg[-\cos f \mathcal{R} + \frac{2 + e\mathrm{cos}\bar{f}}{1 + e\cos\bar{f}}\sin\bar{f} \mathcal{S}\\
- e\cot\iota \frac{\sin(\omega +\bar{f} )}{1 + e\cos\bar{f}}\mathcal{W}\Bigg].
\end{split}
\eeq 
The relevant orbital elements here are the eccentricity $e$, the inclination $\iota$, the nodal angle $\Omega$, the pericenter angle $\omega$ and the semilatus rectum $p$, defined by $p = a(1-e^2)$ where $a$ is the semi-major axis. Then, $\phi$ is the orbital phase as measured from the ascending node, and $\bar{f}$ is the true anomaly, defined by $\bar f \equiv \phi - \omega$. 
Here the noncommutative correction to the radial, cross-track and out-of-plane components of the perturbing acceleration, $\mathcal{R}, \mathcal{S},$ and $\mathcal{W}$ are given by
\beq 
\mathcal{R}_\NC = -\frac{9M^3(1-2\nu)\Lambda^2}{8r^4}\left((\textbf{n}\cdot\boldsymbol{\theta})^2 - \frac{1}{3}\right),
\eeq 
\beq 
\mathcal{S}_\NC = \frac{3M^3(1-2\nu)\Lambda^2}{4r^4} (\bs{\lambda}\cdot\bs{\theta})(\bs{n}\cdot\bs{\theta}),
\eeq 
\beq 
\mathcal{W}_\NC = \frac{3M^3(1-2\nu)\Lambda^2}{4r^4}(\hat{\bs{h}}\cdot\bs{\theta})(\bs{n}\cdot\bs{\theta}),
\eeq 
where $\bs{\lambda}$ is defined as $\partial \bs{n}/\partial\phi$ and $\hat{\bs{h}} = \bs{n}\times \bs{\lambda}$. Expanding out these expressions in Cartesian coordinates in the equations of motion yields a complicated expression that can be further simplified as in \cite{Will:2018ont} by introducing the variables 
\beq 
\bs{e}_P \equiv \bs{n}|_{\phi = \omega} = \bs{e}_\Omega \cos\omega + \bs{e}_\perp \sin\omega,
\eeq 
\beq 
\bs{e}_Q \equiv \bs{\lambda}|_{\phi = \omega} = -\bs{e}_\Omega \sin\omega + \bs{e}_\perp \cos\omega,
\eeq 
\beq 
\hat{\bs{h}} \equiv \bs{e}_P \times \bs{e}_Q = \bs{e}_\Omega \times \bs{e}_\perp . 
\eeq
Here, $\bs{e}_P$ is a unit vector pointing towards the pericenter and $\bs{e}_Q = \hat{\bs{h}} \times \bs{e}_P$. $\bs{e}_\Omega$ is a unit vector which points along the ascending node, and $\bs{e}_\perp = \hat{\bs{h}}\times \bs{e}_\Omega$. $\bs{n}$ and $\bs{\lambda}$ can be analogously translated into the $P$, $Q$, and $h$ coordinates. 

Next, we integrate  Eq. \eqref{eq:dwdt}  from 0 to $2\pi$ to find the noncommutative correction to $\Delta\omega$. We find
\begin{align}
\Delta\omega_\NC =& -\frac{3\pi M^2\Lambda^2(1-2\nu)}{8p}[2 - 3\theta_P^2 - 3\theta_Q^2 + 2(\hat{\bs{L}}\cdot\bs{\theta})\cot \iota] \nonumber \\
&\times(\theta_p\cos\omega + \theta_Q\sin\omega). 
\end{align}
In this expression there is both explicit $\omega$ dependence, as well as implicit $\omega$ dependence in $\theta_p$ and $\theta_Q$. Thus, it is more enlightening to express everything in terms of $\bs{e}_\Omega$ and $\bs e_\perp$. We can then expand $\omega = \omega_0 + \omega^\prime\phi$ and integrate over $\phi$. For $\omega'$, it is sufficient to consider the GR contribution only since $\Delta \omega_\NC$ above is already proportional to $\Lambda^2$. Noting that $\theta_\Omega^2 + \theta_\perp^2 + (\hat{\bs{L}}\cdot\bs{\theta})^2 = 1$ and that for the J0737-3039A/B system, $\iota \approx \pi/2$, we obtain the correction to the pericenter precession as
\beq 
\Delta\omega_\NC = \frac{3\pi M^2\Lambda^2(1-2\nu)}{8p^2}\left[1 - 3(\hat{\bs{L}}\cdot\bs{\theta})^2\right].
\eeq 
 Then, the noncommutative correction to the observable quantity of pericenter precession $\dot{\omega}$, which can be found by dividing $\Delta \omega$ by the orbital period, $P_b$ is
\beq 
\dot{\omega}_{\NC} = \frac{3}{16} \frac{M^{4/3}}{(1-e^2)^{2}}\left(\frac{P_b}{2\pi}\right)^{-7/3}\Lambda^2(1-2\nu)\left[1 - 3(\hat{\bs{L}}\cdot\bs{\theta})^2\right],
\eeq 
where we have used Kepler's law to write $p^2$ in terms of the orbital period.
 Adding this to the GR expression for $\dot{\omega}$ \cite{Stairs:2003eg} we obtain 
 \begin{align}
 \label{eq:omega_dot}
 \dot{\omega} = & 3\left(\frac{P_b}{2\pi}\right)^{-5/3}\frac{M^{2/3}}{1-e^2} \left\{1+ \frac{1}{16} \frac{1}{1-e^2}\left(\frac{P_b}{2\pi M}\right)^{-2/3}\right. \nonumber \\ 
 &\left.  \times \Lambda^2(1-2\nu)\left[1 - 3(\hat{\bs{L}}\cdot\bs{\theta})^2\right] \right\}.
 \end{align}
For completeness, we present the noncommutative corrections to other orbital elements in Appendix \ref{app}.

\subsection{Bounds on $\sqrt{\Lambda}$}

We now derive constraints on $\sqrt{\Lambda}$ with the double pulsar system PSR J0737-3039A/B. We wish to use $\dot\omega$ to constrain the theory. To do so, we need to determine the masses from other observables. Here, we use the Shapiro delay $s$, masss ratio $R$, and the mass functions $f_A$. The noncommutative correction to these observables enter through the Kepler's law at 2PN or higher (see \cite{Sampson:2013wia} for a similar analysis when the metric is modified at 1PN order within the parameterized PN formalism), while the one in $\dot \omega$ in Eq. \eqref{eq:omega_dot} enters at 1PN order. This justifies us to use the GR expressions for $s$, $R$ and $f_A$ to determine the masses and use $\dot \omega$ to test the noncommutative gravity.

\begin{figure} 
    \centering
    \includegraphics[width=\linewidth]{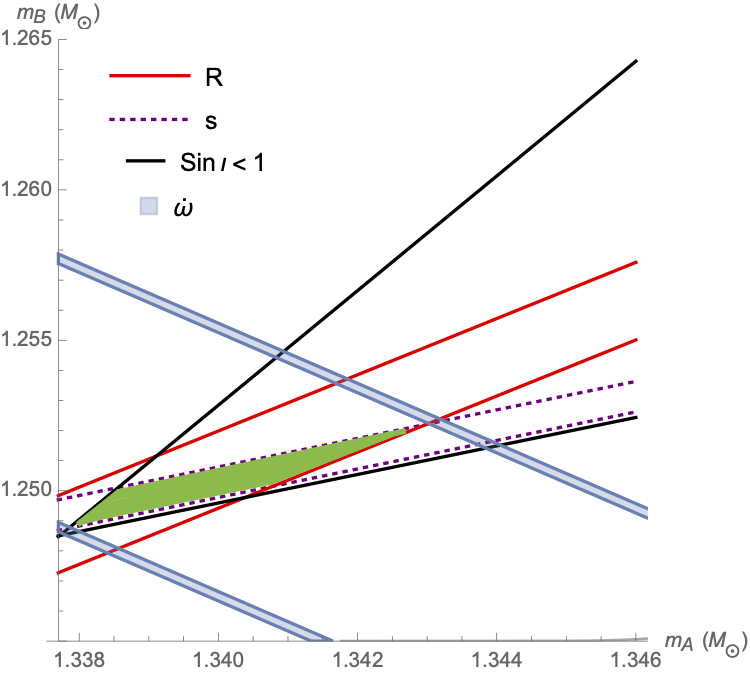}
    \caption{
    Testing noncommutative gravity with the double pulsar binary. The two masses are determined from the mass ratio $R$, the Shapiro delay parameter $s$ and $\sin \iota < 1$ using the GR expressions (since the noncommutative corrections to these observables enter at higher PN orders than that for $\dot \omega$), with the allowed region shown by the green shade. We then vary the noncommutative parameter $\Lambda$ in $\dot \omega$ such that it is consistent with the green shaded region to determined the bound on $\Lambda$. The thickness of $\dot \omega$ in blue corresponds to the measurement error on $\dot \omega$. }
    \label{fig:NSConstraint}
\end{figure}

\begin{figure}
    \centering
    \includegraphics[width=\linewidth]{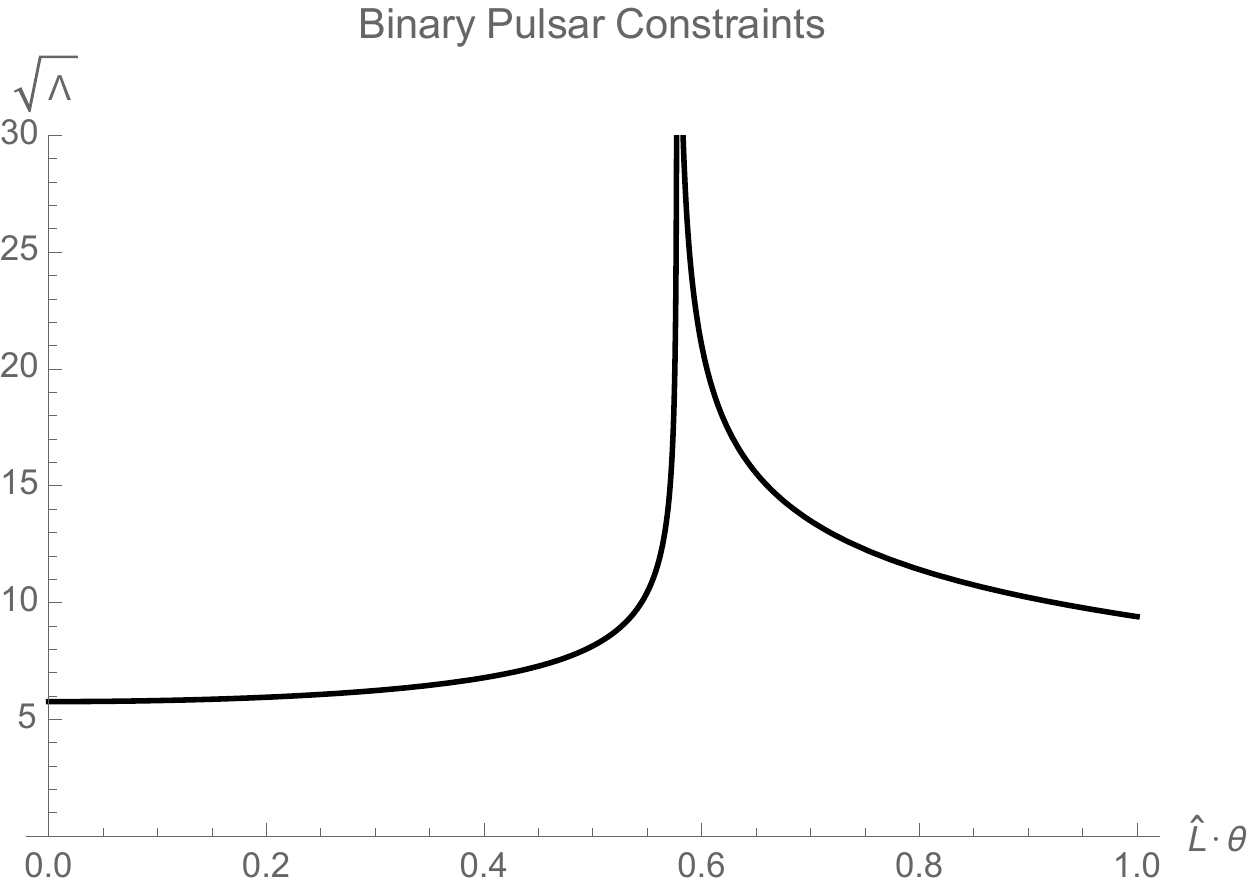}
    \caption{Bounds on $\sqrt{\Lambda}$ as a function of $(\hat{\bs{L}}\cdot\bs{\theta})$ for the double pulsar binary.}
    \label{fig:NSSqrtLambda}
\end{figure}

Figure \ref{fig:NSConstraint} shows these observables plotted as a function of the pulsar masses. The GR expressions for R, s, and the region for which $\sin \iota$ (obtained from the mass function measurements) is less than one are plotted. The overlapping shaded region corresponds to the allowed mass parameter space from these measurements. Any correction to $\dot{\omega}$ must remain within the region of overlap. The upper and lower bounds for $\dot{\omega}$ correspond to variations in the expression  $\Lambda^2\left[1 - 3 (\hat{\bs{L}}\cdot\bs{\theta})^2\right]$ such that the $\dot{\omega}$ curve marginally passes through the overlapping region. 
The thickness in each of the curves corresponds to the uncertainty in the $\dot{\omega}$ measurement and the two curves correspond to the upper and lower bounds on $\Lambda^2\left[1 - 3 (\hat{\bs{L}}\cdot\bs{\theta})^2\right]$. We find that the acceptable range for the noncommutative contribution is 
\beq 
-15600 \lesssim \Lambda^2\left[1 - 3 (\hat{\bs{L}}\cdot\bs{\theta})^2\right] \lesssim 1100.
\eeq 
Then, as we did in the gravitational wave analysis, taking this upper and lower bound, we can  plot $\sqrt{\Lambda}$ as a function of $(\hat{\bs{L}}\cdot\bs{\theta})$ as in Fig. \ref{fig:NSSqrtLambda}.
We can see that there is again a particular value of $(\hat{\bs{L}}\cdot\bs{\theta}) = \sqrt{1/3}$ that we are not able to place a constraint as was the case for the GW analysis, however we are still able to place bounds for the rest of the range. 
We find that the binary pulsar bounds are actually less stringent than those found from the gravitational wave events by approximately an order of magnitude. However, these constraints remain consistent with the general statement that the noncommutativity parameter must be of order unity. 

\section{Conclusion}\label{V}

We have explored noncommutative gravity in light of observations from LVC gravitational wave events as well as the binary pulsar system J0737-3039A/B. We have focused on the lowest order noncommutative effects entering at 2PN in the binary system acceleration. The time component of the noncommutative tensor, $\theta^{0i}$ enters as a 2PN correction to the acceleration. When this effect is propagated through, we find that there is a phase shift in the gravitational  waveform again entering at 2PN, shifting the $\varphi_4$ coefficient. Similarly, in the case of binary pulsar dynamics, we find that the correction to the acceleration leads to a noncommutative contribution to the pericenter precession.  

An updated and more rigorous analysis than in previous work has been performed to use gravitational wave events and the binary pulsar system PSR J0737-3039A/B to constrain the space-time component of the noncommutativity tensor. We find that the gravitational wave events including GW151226, GW170608, GW170814 and GW170817 are more constraining than the binary pulsar event PSR J0363-3039A/B by approximately an order of magnitude. However, the more stringent GW constraints are consistent with previous results, findng that the the quantity $\sqrt{\Lambda}$ is constrained to be of order unity. 

A few different avenues exist for future work. For example, it would be interesting to constrain the theory from the preferred frame effect \cite{Will:2018ont}.
It would also be of interest to investigate the effects of the spatial component of the noncommutative tensor, $\theta^{ij}$, which enters at 3PN and has potential implications for e.g. string theory. Additionally, it would be valuable to explore the model dependence of the effects that we have discussed, and work towards a more general understanding of how noncommutative gravity may come into play with these observables.

\acknowledgments
 K.Y. acknowledges support from NSF Award PHY-1806776, NASA Grant 80NSSC20K0523, a Sloan Foundation Research Fellowship and the Ed Owens Fund. 
K.Y. would like to also acknowledge support by the COST Action GWverse CA16104 and JSPS KAKENHI Grants No. JP17H06358.

\appendix
\section{Noncommutative Corrections to Osculating Orbits}
\label{app}
In addition to the noncommutative correction to the pericenter precession, $\dot{\omega}$, the noncommutative correction to the acceleration also induces corrections to the other orbital parameters, $p, e, i$, and $\Omega$, described in Section \ref{IV}. The  ``Lagrange planetary equations" for these quantities are \cite{Will:2018bme}
\allowdisplaybreaks
\begin{align}
    \frac{dp}{dt} =& 2\sqrt{\frac{p^3}{M}}\frac{\mathcal{S}}{1 + e\cos\bar{f}},\\
    \frac{de}{dt} =& \sqrt{\frac{p}{M}}\left[ \sin\bar{f} \mathcal{R} + \frac{2\cos\bar{f} + e + e \cos^2\bar{f}}{1 + e\cos\bar{f}}\mathcal{S}\right],\\
    \frac{d\iota}{dt} =& \sqrt{\frac{p}{M}}\mathcal{W}\left(\frac{r}{p}\right) \cos\phi, \\
    \frac{d\Omega}{dt} =& \sqrt{\frac{p}{M}} \mathcal{W} \left(\frac{r}{p}\right)\frac{\sin\theta}{\sin \iota}.
\end{align}
Plugging in the expressions for $\mathcal{S}, \mathcal{W}$, and $\mathcal{R}$ obtained in Section \ref{IV}, it is straightforward to obtain 
\allowdisplaybreaks
\begin{align}
\Delta p_\NC =& 0,\\
\Delta e_\NC =& 0, \\
\Delta \iota_\NC =& \frac{3\pi M^2 \Lambda^2(1-2\nu)}{4p}(\hat{\bs{L}}\cdot\bs{\theta})(\theta_p\cos\omega - \theta_Q\sin\omega),\\
\Delta \Omega_\NC =& \frac{3\pi M^2\Lambda^2(1-2\nu)}{4p}(\hat{\bs{L}}\cdot\bs{\theta})(\theta_p\cos\omega + \theta_Q\sin\omega)\csc i. \nonumber \\
\end{align} 
As in $\Delta\omega$ there is both explicit and implicit $\omega$ dependence in these expressions. Using the same expansion method, we obtain for the noncommutative contributions to the orbital parameters:
\begin{align}
 \Delta p_\NC =& 0,\\
 \Delta e_\NC =& 0,\\
 \Delta \iota_\NC =& \frac{3\pi M^2\Lambda^2(1-2\nu)}{4p^2}(\hat{\bs{L}}\cdot\bs{\theta})\theta_\Omega,\\ 
\nonumber \\
 \Delta \Omega_\NC =& \frac{3\pi M^2\Lambda^2(1-2\nu)}{4p^2}(\hat{\bs{L}}\cdot\bs{\theta})\theta_\perp \csc i.
\end{align}

\bibliography{ref}{}
\end{document}